\documentstyle [12pt] {article}

\begin{document}
\title {\large \bf  Representations\\ of The Coordinate Ring of
$ GL_{q}(3) $}
\author {Vahid Karimipour}
\date { }
\maketitle
\begin {center}
{\it  Department of Physics , Sharif University of Technology\\
P.O.Box 11365-9161 Tehran, Iran\\
Institute for studies in Theoretical Physics and Mathematics
\\ P.O.Box 19395-1795 Tehran, Iran \\
Email. Vahidka@irearn.bitnet}.
\end{center}
\vspace {10 mm}
\begin {abstract}
{It is shown that the finite dimensional ireducible representations of the
quantum
matrix algebra $ M_q(3) $ ( the coordinate ring of $ GL_q(3) $ ) exist only
when q is a root of unity
( $ q^p  = 1 $ ). The dimensions of these representations can only be one
 of the
following values: $ p^3 \ ,  \ { p^3 \over 2 } \ , \ { p^3 \over 4 }
\ , \ $ or
$ \ { p^3 \over 8 } $ .
The topology of the space of states ranges between two extremes , from a
3-dimensional
torus $ S^1 \times S^1 \times S^1 $ ( which may be thought of as a
generalization
of the cyclic representation ) to a 3-dimensional cube $ [ 0 , 1 ]\times
[ 0 , 1 ]\times   [ 0 , 1 ] $ .}
\end{abstract}
\noindent
\newpage
{\large \bf I. Introduction}\\

Currently there is a rather well developed theory of representation
[1-5] for quantized universal enveloping algebras [6-10].
However the representations of the dual objects, that is the quantizatio
n of the
algebra of functions on the group , or their coordinate rings, has not been
studied
systematically, except for the  case of $ GL_q (2) $\ \  [11].
In this paper we  extend these studies to the next simplest case , that is
the coordinate ring of $ GL_q(3) $ .
Why such a problem is interesting becomes clear when one notes that in the
case of
a quantised enveloping algebra $ U_q(g) $, one has a rather natural
decomposition of the generators of $ U_q(g)$
into the set of Cartan generators , q-anoloug of simple roots , and
q-analogue
of
positive non-simple roots which are defined via q-adjoint action [2].
Thus in any
$ U_q ( g ) $ module $ V $ one needs to define only the action of simple roots
le roots
on
$ V $. For the case of quantum matrix groups , such a decomposition does
not
exist.
What we will do in this paper is to give such a decomposiyion  for
$ M_q (3) $
and then
construct its finite dimensional representations. These representations
exist only
when q is a root of unity and depend on a number of parameters which
determine the
topology of the space of states. Depending on these parameters the
topology can
range between two extremes, from $ S^1 \times S^1 \times S^1 $\ \ \
 to \ \ \ $
[0 , 1] \times
[0 , 1] \times  [0 , 1] $.

The structure of this paper is as follows : In section \ II) \ we briefly
discuss the
structure of $ M_q(3) $ . In section\ III)\  we introduce a particular
subalgebra of
$M_q(3) $\  and construct its finite dimensional representations in s
ection IV).
We then use this subalgebra to construct the representations of $
M_q(3) $ in
section
$ V $. At the end of the paper we prove that the only finite dimensional
irreducible
representations of $ M_q (3 ) $ are of dimension $  p^3 $ ,
$ {p^3\over 2 } $ ,
$ {p^3\over 4 } $ or ${p^3\over 8 }$ .
\vskip 1cm
{\bf II. The Structure of $ GL_q(3) $}

The quantum  Matrix algebra $ M_q(3)$ is genereated by the elements of a
matrix
\begin{equation} T = \left( \begin{array}{lll} a & b & c   \\ d & e & f
\\ g & h & k  \end{array} \right)\end{equation}
subject to the relations [8]:
\begin{equation} R\ T_1 T_2 = T_2 T_1 R \end{equation}
where R is the solution of the Yang Baxter equation corresponding to
$ SL_q (3)
$[7].
The relations
obtained from (2) can be expressed neatly in the following form :
For any $ 2\times 2 $ submatrix ( i.e : like the one formed by the elements
$ b , c, e, $ and $ f $ )
 the following relations hold:
$$ b c = q\  cb \hskip 1cm ef = q\  fe  $$
\begin{equation} be = q\  eb \hskip 1cm cf = q\  fc \end{equation}
$$ ec = ce \hskip 1cm bf - fb = (q - q ^{-1}) \ ce $$

One can also prove the following  more general
type of relations \begin{equation} bf^n - f^n b = q^{-1} ( q^{2n}
-1 ) f^{n-1} ce \end{equation}
$$  fb^n - b^n f = q ( q^{-2n} -1 ) b^{n-1} ce $$
by induction from (3).
{\bf Remark}:\\
These relations are only a small part of the relations
obtained
from (2).All the other relations can be simply read by looking at other
submatrices.( i.e: $ df = q fd\ \  , dg = q gd\ \  , dk - kd =
( q - q^{-1} )
fg\ \  , $  etc
). Hereafter when we refer to ( 3,4 ) we mean all the relations of which
(3 and 4 ) are
 samples.
Thus this algebra has  many $ GL_q(2) $ subalgebras ( i.e: the set of
elements
$ a,\  c,\  d,\  and \ \ f $ or $ d \ \ f \ \ g \ \  $ and $ k $
generate two such  $ GL_q(2) $ subalgebra ). Obviously these are not
Hopf subalgebras.

$ GL_q(3) $ has also a quantum determinant D which is central:
\begin{equation} D = a \Delta_a - q b \Delta_b + q^2  c \Delta _c \end
{equation}
where $ \Delta_a \ \ \ \Delta_b \ \  $ \ \ and \ \ $ \Delta_c $\ \  are the
quantum cofactors of the elements\ \  $ a \  \ b  $ \ \ and\ \  $ c $\
respectively :
\begin{equation} \Delta_a = ek - q fh \hskip 2cm \Delta_b = dk - q fg \
hskip 2cm \Delta_c =
dh - q eg \end{equation}
One can also see from (3) that the elements $ c \ \  e $  and $ g $
commute with each other .
\vskip 1cm
{\large \bf III. A Good Subalgebra of  $ GL_q ( 3 )$ }\\

It is much simpler to construct first the representations of a
subalgebra of $ M_q (3 ) $.
This subalgebra has the nice feature that while its commutation
relations are much
simpler than those of\ \  $ M_q(3) $\ \  one can use its
representations in a very
direct way to construct\  $ M_q (3) $ \ representations.
In this section we introduce this subalgebra which we will denote by $ A $
and study
in some detail its structure . This subalgebra is  generated by
all the elements of the matrix $ T $  ( except  $ a $  and  $ k $ )
plus two quantum
cofactors $ \Delta_k $ , $ \Delta_ a  $ and the quantum determinant $ D $ .
Hereafter we denote $ \Delta_a $ and $ \Delta_k $ by $ \Delta $ and
$ \Delta ' $
respectively .
Clearly the element $ \Delta $  being the q-determinant of the submatrix
$  \left( \begin{array}{ll} e & f  \\ h & k \end{array} \right) $
commutes with
the elements $ e, f, h $  and  $ k $ . A similar statement holds true for
$ \Delta' $ . ( i.e:\ \ $ \Delta' $ commutes
with $ a , b, d $ and $ e $  ) .
Using (3) its straightforward to verify the following commutation
relations:
$$ b \Delta  = q \Delta b \hskip 3 cm c \Delta  = q \Delta c $$
\begin{equation} d \Delta  = q \Delta d \hskip 3 cm
g\Delta  = q \Delta g \end{equation}
and
$$ c \Delta'  = q^{-1} \Delta' c \hskip 3 cm g \Delta'  = q^{-1}
\Delta' g $$
\begin{equation} f \Delta'  = q^{-1} \Delta' f \hskip 3 cm h \Delta'
= q^{-1}  \Delta' h
\end{equation}
Consider an A-Module V. The above relations in conjunction with (3) show
that if
by $ \vert \lambda , \mu,  \nu > $ we denotes  common eigenvectors of
$ c ,
e , $ and $ g $  with eigenvalues $ \lambda , \mu $ and $ \nu $
 respectively,
then :
$$ f\vert \lambda , \mu , \nu > \ \ \propto \ \  \vert q \lambda , q \mu ,
\nu > \hskip 1cm  ( f\ \ , q\ \  \longrightarrow b\ \  ,\ \  q^{-1} )  $$
\begin{equation} h\vert \lambda , \mu,  \nu > \ \ \propto\ \  \vert
\lambda , q \mu ,
q \nu > \hskip 1cm ( h\ \ , q\ \  \longrightarrow d\ \  ,\ \  q^{-1} )
\end{equation}
$$ \Delta \vert \lambda , \mu \nu >\ \  \propto \ \ \vert q \lambda ,  \mu,
q\nu > \hskip 1cm  ( \Delta \ \ , q\ \  \longrightarrow \Delta'\ \  ,\ \
q^{-1} )  $$
Therefore in any A-module V the elements\  $ f\  , \ h , $ \ \ and \ \  $
\Delta $ ( resp.  $ b\  , \ d , $ \ and \ $ \Delta' $ ) act as raising
,(resp. lowering ) operators in the directions of eigenvalues of $ ( c,e )$
$ ( e,g ) $ and $ ( c,g ) $ respectively.
Note that the states \ $ a \vert \lambda,  \mu,  \nu > $\ \  and \ \  $ k
\vert \lambda , \mu,  \nu > $\  are no longer eigenstates of $ c , e, $
 and $g$.
Stressing the analogy
with representations of quantum algebras , we note that the operator
$ \Delta $
rather than $ k $ , play the role of q-analouge of positive root. The
q-analouge of
simple roots being $ f ,$ \ and \ $ h ,$.
The following lemma is needed for construction of A-modules.

{\bf Lemma 1}:

i)$ bf^l = f^l b + q^{-1}(q^{2l}-1)f^{l-1}ec $

ii)$ dh^m = h^m d + q^{-1}(q^{2m}-1)h^{m-1}eg $

iii) $ a \Delta = q^2 \Delta a + ( 1-q^2 ) D  $

iv)$ \Delta'  \Delta = q^2 \Delta \Delta'  + ( 1-q^2 ) De $

Where $ D $ is the quantum determinant of the matrix T ( see (5) ) .

{\bf Proof }\\:
Only iii) and iv) need proofs.
For iii) we note that passing $ a $ through $ \Delta $  and using the
commutation relations (3) we find
$$ a \Delta = \Delta a + ( q - q^{-1} ) ( ecg + bdk - q fbg - q cdh ) $$
from (3) we have:
$$ bdk - q fbg = bdk - q ( bf - ( q-q^{-1})ce ) g = b\Delta_b +
( q^2 - 1 ) ceg $$
therefore the sum of the terms in the bracket is equal to $ b \Delta_b -q c
\Delta_c $
and the above
equation is transformed to the following form:
$$ a\Delta = \Delta a + ( q-q^{-1}) ( b \Delta _b - q c \Delta _c ) $$
By using the expression of the quantum determinant (5) we finally arrive at
(iii).
The proof of iv) is obtained by induction from iii).  $ \hskip 2cm $ QED.

{\bf Coralarry}:

i)$ a \Delta^n = q^{2n } \Delta^n a + ( 1-q^{2n}) D \Delta^{n-1}  $

ii) $ \Delta ' \Delta^n = q^{2n} \Delta^n \Delta' + ( 1-q^{2n}) D e \Delta^
{n-1}   $

These formulas are proved by induction from (10) and (11) .
\vskip 1cm
{\large \bf IV. Finite Dimensional Representations of A }\\
We call those $ A$ modules in which one or more of the generators identica
lly
vanish,
trivial modules . In this cases the representation reduces to that of a
simpler
algebra ( a reduction of $ M_q(3) $ obtained by setting that element
equal to
0 ) Clearly the interesting
representations are nontrivial ones to which we restrict ourselves in the
 rest
of this paper.
We present the condition of nontriviality of $ A $ modules in the
following:
( see also [11] )

{\bf Proposition: }\\

An $ A $ Module V is trivial if one of the following sets contain nonzero
elements:
$$ K_{c} = \{ v\in V \ \ \ \vert c v = 0 \} $$
$$ K_{e} = \{ v\in V \ \ \ \vert e v = 0 \} $$
$$ K_{g} = \{ v\in V \ \ \ \vert g v = 0 \} $$
{\bf Proof:}\\
Without loss of generality , lets assume that \ $ K_c \ne\  \{ 0\}\  $ ,
then since  $ K_c $  is a subspace of V , we can choose a basis for it
like :
$ \{ e_1 , . . .  e_ m \} $ . From the relations (3) we see that \  $ fK_c\
, \ h K_c\  ,
\ \Delta K_c\ , \ g K_c\ , \ e K_c\ , \ b K_c\ , \ d K_c\ $   and $ \Delta'
K_c\  $ are all subspaces of $ K_c $ . Therefore the vectors $ \{ e_i\} $
transform among themselves under the action of A and hence $ K_c$  is an
invariant subspace of V . Since the representation is irreducible $
K_c = V $ .
Therefore
$ c  K_c = c  V = 0 $  and the representation is trivial.$ \hskip 2cm $ QED.

Hereafter we assume that  $ K_c = K_e =  K_g = \{ 0 \} $ .
We first state and prove the following:

{\bf Proposition 4 :}\\
Any nontrivial A-Module is also a nontrivial $ M_q(3) $ module and vice versa.
versa.

{\bf Proof:}\\ Only the first part of the proposition needs a proof. Let $
W $  be
a nontrivial A module. This means  among other things that $ K_e = \{0\} $ .
To prove that W is also an $ M_q(3)$ module we
need to show that the actions of $ a $ and $ k $ are defined on W .
We choose as the basis of W the set of common eigenvectors of \ \ $ c ,
e,\ \
$and $ g $:
$$ B_W = \{ v_i\ \ \ \ \ \   1 \leq i \leq N ,\ \ \ \  c\  v_i = \lambda_i\
v_i ,\ \
\ \ \ e\  v_i = \mu_i\  v_i ,\ \ \ \ \ g\  v_i = \nu_i\  v_i\ \ \ \} $$
 Since W is an A module the action  of $ \Delta $ and $ \Delta '$ on W are
 defined.
Therefore  \begin{equation} \Delta v_i = \sum _j \Delta_{ij} v_j
\end{equation}
$$ \Delta' v_i = \sum _j \Delta'_{ij} v_j $$
since $ \Delta'= ae - q bd $ we conclude from (10-b)that:
\begin{equation}a v_i = { 1\over \mu_i } (\  \sum \Delta'_{ij}\  v_j +
q b d\
 v_i  )\end{equation}
where we have used the fact that $ \mu_i \ne 0 $ due to the nontri
viality of W ,
( proposition 3 ).
We also note that thanks to the commutation relations (3)$ \Delta = ek - q
fh $  has an equivalent description, namely  $ \Delta = ke - q^{-1} fh $ .
Hence we obtain from (10-a) :
\begin{equation} k v_i = { 1\over \mu_i } ( \sum \Delta_{ij} \ v_j +
q^{-1} f h\  v_i )
\end{equation}
Equations (11) and (12) show that the actions of $ a $  and $ k $ are
defined
on W and hence W is an $ M_q ( 3) $ module.$ \hskip 2cm $ QED.

{\bf Proposition 5: }\\

 Finite dimensional representations of  $ A $  exist only when $ q $ is
a root
of unity ( see also [11]).

{\bf Proof: } \\ Suppose that q is not a root of unity .  Let $ v_0 $
be a common eigenvector of $\ \ c \ ,\ e \ \ $ and $ g $ .
$$ c\  v_0 = \lambda \ v_0 \hskip 1.5cm   e\  v_0 = \mu\  v_0
\hskip 1.5cm    g\  v_ 0 = \nu\  v_0  $$
Consider the following string of states: $ v_ l = f^l v_0 $ . One can see
that these states
are eigenvectors of $\  c \ \ e \ $  and  $ g $ .
 $$ c\  v_l = \lambda q^{ l }\  v_l\ \ \ \ \ \  e\  v_l  = \mu q^l\  v_l
 \ \ \ \ \ \  g\  v _l = \nu\  v _l $$ Since all these eigenvalues are
different,
to have finite  dimensional  representations  one must have
$ f^n v_0 = 0$
for some  $ n $
while  all  the  vectors  $ v_l $  for  $ l<n $ are  independent.
 Consider now
the string  of  states   $ b^{l'} u_0 $
where $ u_0 = v_{n-1}\ \ \ .   $ Again there must exist a positive
integer $ n'$
such that  $ b^ {n'} u_0 = 0 $\ \ \  and $ b^{ n'-1} u_0 \ne 0 $.
Then one will
have
$$ 0 = fb^{n'}u_0 = {q}( q^{-2n'}-1) b^{n'-1} ce u_0  = {q}( q^{-2n'}-1)
\lambda \mu q^{2(n-1)} b^{n'-1} u_0  $$
which means that q must be a root of unity.
Hereafter we assume that $ q$  is a root of unity : $ q^p = 1  $ .$
 \hskip 2cm $
QED.

We now construct the finite dimensional representations of $ A $ .
First we note that when $ q^p = 1 $ then all the elements $ f^p \ \ h^p\ \
\Delta^p \ \   b^p \ \ d^p \ \ $\ \   and\ \  $  \Delta'^p $  are c
entral. Therefore
using Schur's Lemma we set them equal to $ \eta_f \ \ \eta_h \ \ \eta_
{\Delta }
\ \ \  \eta_b  \ \  \eta_d  \  \ $ \ \  and   $  \ \ \eta_{ \Delta' }   $
respectively.
As we will see various kinds of representations depend on the values of
 these
parameters.
We denote the vector $ v_0 $  defined in (13)  by $ \vert 0 >\
\equiv \vert 0, 0, 0> $  and consider the
3-dimmensional cube
of states :
\begin{equation} W = \{ \vert l ,m ,n > = f^l h^m \Delta^n \vert 0 >
\ \ \ \ 0\leq l , m ,
n \}\end{equation}
These vectors are all eigenvectors of $ c \ \ g \ \ \ $ and $ e $ .
$$ c\vert l,m,n> = q^{l+n}\lambda  \vert l,m,n> $$
\begin{equation} e\vert l,m,n> = q^{l+m} \mu \vert l,m,n>\end{equation}
$$ g\vert l,m,n> = q^{m+n} \nu \vert l,m,n> $$
We will show that W spans an invariant
submodule of V .
Obviously we have:
$$ f\vert l, m, n > = \vert l+1 , m , n > \hskip 1cm f\vert p-1, m, n > =
\eta_f \vert 0, m, n > $$
\begin{equation} h\vert l, m, n > = \vert l , m+1 , n > \hskip 1cm h
\vert l, p-1, n > =
\eta_h \vert l, 0, n > \end{equation}
$$ \Delta \vert l, m, n > = \vert l , m , n+1 > \hskip 1cm \Delta \vert l,
m, p-1 > = \eta_{\Delta } \vert l, m, 0 > $$
Now we define the action of $ b \ \ d \ \  $ and $ \Delta' $ on
$ \vert 0 > $
as follows:
\begin{equation} b\vert 0 > = \alpha_0 \vert p-1 , 0 , 0 > \ \ \ \
\ d\vert 0 > = \beta_0
\vert 0 , p-1 , 0 >   \ \ \ \ \  \Delta '\vert 0 > = \gamma_0
\vert 0 , 0 , r-1>
\end{equation}
Then we have:

{\bf Lemma :}\\

i)$ b\vert 0, m, n > =q^{m+n}\alpha_0  \vert p-1 , m , n > $

ii)$ d\vert l, 0, n > = q^{ l+n} \beta_0 \vert l , p-1 , n  > $

iii)$  \Delta' \vert l, m, 0 > = q^{l+m}\gamma_0  \vert l , m , p-1 >  $

{\bf Proof :}\\ Using (3,17) the verification is straightforward.

{\bf Lemma : }

i)$ b\vert l, m, n > =q^{m+n}( \alpha_0 \eta_f + q^{-1} ( q^{2l}-1)\
lambda \mu )
\vert l-1 , m , n > $

ii)$ d\vert l, m, n > = q^{ l+n}( \beta_0 \eta_h + q^{-1} (
q^{2m}-1)\mu \nu )
\vert l , m-1 , n >$

iii)$  \Delta' \vert l, m, n > = q^{l+m} ( q^{2n} \gamma_0 \eta_{\Delta}+
( 1-q^{2n} ) \mu \eta ) \vert l , m , n-1 > $

{\bf Proof :} We only prove part i) . The other two parts are similarly
verified. Passing b through $ f^l $ and using part i) of Lemma 1 we have :
$$ b \vert l, m, n> =bf^lh^m\Delta^n \vert 0 > = ( f^l b + q^{-1}
( q^{2l} -1)
ec) h^m \Delta^n \vert 0 > $$
$$ = f^l b \vert 0, m, n > + q^{-1} ( q^{2l}-1) q^{m+n}\lambda \mu
\vert 0, m, n > $$
Using part i) of the previous lemma we arrive at the desired result:
$$ b\vert l, m, n> = q^{n+m} f^lh^m \Delta^n b\vert 0 > + q^{-1}
 ( q^{2l}-1) q^{n+m} f^{l-1}h^m\Delta^n \lambda \mu \vert 0 > $$
\vskip 1cm
{\large \bf V. Finite Dimensional Representations of
  $ M_q(3) $}\\

We now use proposition 4 and define the actions of $ a $ and $ k
$ on $ W $ .
Thanks to the commutation relations (3) and the
definition of states (14) one only requires to find the action of $ a $
and $ k $  on the
state $ \vert 0 > \equiv \vert 0, 0, 0 > $ .
To calculate  $ k\ \vert 0 > $ we note that $ \Delta $  has an equivalent
expression namely $ \Delta = ke-q^{-1}fh $.
Therefore $$ \vert 0, 0, 1> = \Delta \vert 0 > = ( ke-q^{-1} fh ) \vert 0 >
= \mu k\vert 0 > - q^{-1} \vert 1, 1, 0 > $$
from which we find :
\begin{equation} k \vert 0> = {1\over \mu } ( \vert 0, 0, 1 > + q^{-1}
\vert 1, 1, 0 > )
\end{equation}
Similarly we note that :
$$ \Delta' \vert 0 > = \gamma_0 \vert 0, 0, r-1 > = ( ae - q bd ) \vert 0 >
= \mu a\vert 0 > - q  \alpha_0 \beta_0 \vert r-1 , r-1 , 0 >  $$
from which we find : \begin{equation} a\vert 0 > =  \gamma _0 \vert
0, 0, r-1 > + \alpha_0
\beta_0 \vert r-1, r-1, 0 >\end{equation}
Using the commtation relations one readily find the action of
 $ a$ and $ k $
on any other state $ \vert l, m, n > $ and can verify
that equations ( 15-19) toghether with lemmas 6 and 7 define a finite
dimenstional representation of $ M_q (3)$ .
The action of the operators is shown in fig.(1). The operators $ f, h, $
and $  \Delta $  ( $ b , d , $ and $ \Delta' $ )
act as raising ( Lowering ) operators in the directions
$ l, m, $ and $ n $ respectively .
Each state in this cube is common eigenvector of $ c , e $ and $ g $ .

Note that the parameters $ \alpha_0 , \beta_0 , $  and $ \gamma_0 $are not
independent of the parameters  $ \eta_f ,\eta_h  $ and $ \eta_{\Delta'} $ .
Their relation
is given in the appendix.

We conclude this paper with a proposition which contains our main result.

{\bf Proposition 8 :}\\

The only finite dimensional irreducible representations of $ M_q
( 3 ) $  are of dimension $ p^3 $ ( when p is odd ) and dimensions
$ p^3 $ ,
$ {p^3\over 2 } $ , $ {p^3\over 4 } $ or ${p^3\over 8 }$ ( When $ p $ is
 even ).

{\bf Proof: }\\  Our style of proof is a generalization of the one given in
[11] for the case of $ M_q(2) $.
Let V be an $ M_q ( 3 ) $ module with dimesion d . First consider the case
where none of
the pair of parameters\ \   $( \eta _f , \eta _b  ) ,\ \  ( \eta _h
\eta _d )
\ \   or\ \  ( \eta _{ \Delta } ,  \eta _ {\Delta'} ) $ \ \ are zero .

In this case $ d $ can not be greater than $ p^3 $, otherwise the cube
W ( see figure 1 ) will span an invariant submodule
which contradicts the irreducibility of V . The dimension of V can
not be less than $ p^3 $  either
since this means that the lenght of one of the sides of the cube W
( say in the l-direction)
must be less than p . Therefore there must exist a positive integer $ r<p
$ such that $
f^r \vert 0,m,n> = 0 $ which means that $ \eta_f \vert 0,m,n> = f^{p-r}f^r
\vert 0,m,n> = 0 $ )
contradicting the original assumption.
The topology of the space of states in this case is a 3 dimesional torus
( $ S^1 \times S^1 \times S^1 $ )

Now consider the case when one of the factors\ \  $ \eta_f \eta_b ,
 \ \  \eta_h
\eta_d  \ \ $ or $ \ \  \eta_{ \Delta} \eta_{ \Delta' } $\ \  say the first
one is zero. There are two subcases to consider:

a) $ \eta_f = \eta_b = 0 $

If\ \  $ d <  P^3 $\   there must exists an integer like $ r<p$ such that
$ f^r \vert 0,m,n> = 0 $ and $ f^l\vert 0,m,n> \ne 0 $ for
 $ l\ < \ r  $. Now denote $ f^r\vert 0,m,n> $ by $ u_0 $ and consider the
 string of states $ b^{l'}u_0 $ .This string of states ( for
fixed m and n ) must terminate somewhere. Thats there must exists
 an integer
like $ r'$ such that $ b^{r'}u_0 = 0 $ and $ b^{r'-1}u_0 \ne 0 $
Therefore $$ 0 = fb^{r'}u_0 = ( b^{r'} f + q ( q^{-2r'}-1)b^{r'-1}ce )
u_0 =
q(q^{-2r'}-1)\lambda \mu q^{ 2l + m + n } b^{r'-1}u_0 $$
which means that $ q^{2r'}=1 $ or $ r'={p\over 2 } $ . r' is in fact the
 lenght
of the edge of the cube W in the l direction , the other two edges being of
lenght p . The dimension of V is in this case $ { p^3\over 2 } $ . The
topology
of the space of states is in this case
$ [0 , 1]\times S^1 \times S^1 $. One can easily convince himself that
in this
case irreducible representations with dimension
greater than $ p^3 $ do not exist , but there exist irreducible
representations
with dimension equal to $ p^3 $ in which:
$$  f \vert p-1, m , n > = 0 \hskip 2cm b \vert 0, m, n, > = 0  $$

b) $ \eta_f = 0 $ and $ \eta_b \ne 0 $ or vice versa.

This case can arise only when p is odd. Since for even p the vanishing of
$ \eta_f $ implies the vanishing of $ \eta_b $ . In fact :
from (A-1) we have $ \eta_b \ =  \prod_{i=1}^{p-1} ( q^{2i}-1 ) = 0 $
For\ \  $ \eta_f = 0 $\  and \ $ \eta_b \ne 0 $ we have:
$$ f\vert p-1, m, n > =  0 \hskip 2cm b\vert 0, m, n > = q^{m+n} \alpha_0
\vert p-1 , m , n > $$
In this case the operator b traverses the full circle of states
$ \vert l,m,
n > $ ( for fixed m and n )
while f does not, and the representation is $ p^3$ dimensional.The topology
is still a 3 dimensional torus.

Repeating the above reasoning for the other cases where the other parameters
$\ \  (\eta_h
\eta_d )\ \  $ and/or $\ \ (\eta_{\Delta} \eta_{\Delta'} )\ \ $
vanish completes the proof of this proposition and verifies the statement
claimed in the abstract
conserning the topologies of the representations.
\vskip 1cm
{\bf Acknowledgement }: The author whishes to thank M. Khorami
, A. Aghamohammadi , and S. Shariati for enlightening discussions.

\vskip 1cm

{\bf Appendix A } : Relation Between The Parameters

i) The following relations exist between the parameters $ \alpha_0
 , \beta_0
\ \ $ and $ \gamma_0\ $ \ and the other parrameters of the representation.
\begin{equation} \eta_b = \alpha_0 \prod_{i=1}^{p-1} x_i  \hskip 2cm
 \eta_d = \beta_0 \prod_{i=1}^{p-1} y_i \hskip 2cm
 \eta_{ \Delta' } = \gamma _0 \prod_{i=1}^{p-1} z_i \end{equation}
Where $$ x_i = \alpha_0 \eta_f + q^{-1} \lambda \mu ( q^{2i}-1) $$
$$ y_i = \beta_0 \eta_h + q^{-1} \mu \nu ( q^{2i}-1)  $$
$$z_i = q^{2i} \gamma_0 \eta_{\Delta} +\mu \eta (1- q^{2i})  $$

We only prove the first relation , the other two relations are
similar.
Repeated application of part (i) of lemma 7 gives:
$$ b^{p-1} \vert p-1, m, n > = ( q^{m+n} )^{-1} \prod_{i=1}^{p-1} x_i \vert
0, m, n > $$
Acting on both sides with $ b $ and using part ( i ) of lemma 6  and
the fact
that $ b^p = \eta_b $
proves the assertion.

ii) The value of the parameter $ \eta $  can be determined by acting
on any state
$ \vert l, m, n > $ with
D . After a straightforward calculation one obtains :
\begin{equation}  \eta = -q^{-3} \lambda \mu \nu  + ( {\gamma_0 \eta
_{\Delta}\over \mu }
-{\alpha_0\beta_0  \eta_{h}\eta_f \over q \mu }
+{ \nu \alpha_0 \eta_{f}\over q^2 }+  {\lambda\beta_0 \eta_{h}\over q^2 })
\end{equation}

\vfil\break

{\large \bf References: }\\
\begin{enumerate}
\item {1-} G. Lusztig , Adv. Math. 70, 237 (1988); Contemp. Math.
82,59 (1989)
\item {2-} M. Rosso , Commun. Math. Phsy. 117, 581 (1988) ; 124, 307 (1989)
\item {3-} R. P. Roche and D. Arnaudon , Lett. Math. Phsy. 17, 295 (1989)
\item {4-} C. De Concini and V. G. Kac , Preprint (1990)
\item {5-} P. Sun and M. L. Ge , J. Phys. A 24, 3731 (1991)
\item {6-} V. G. Drinfeld ; Proceeding of the ICM ( Berkeley, Berkeley , CA,
1986) p.798
\item {7-} M. Jimbo ; Lett. Math. Phsy. 10, 63 ( 1985) ; 11,247 (1986)
\item {8-} N. Reshetikhin , L. Takhtajan , and L. Faddeev ; Alg. Anal. 1, 1
78 (
1989) in Russian
\item {9-} S. Woronowics , Commun. Math. Phys. 111, 613 (1987); 130,
 387(1990)
\item {10-} Y. Manin CRM Preprint ( 1988)
\item {11-} M. L. Ge, X. F. Liu , and C. P. Sun : J. Math. Phys. 38 (7) 1992
\end{enumerate}
\vfil\break

FIGURE CAPTION

Figure  1: The states of the A module  W .Depending on the value of the
parameters
$ \ \ \eta_f . . .\eta_{\Delta} \ \ $ oposite sides of this cube can be
 identified
independently.
\end{document}